\documentstyle[preprint,aps,eqsecnum]{revtex}

\def\beq#1{\begin{equation} \label{#1}}
\def\eeq{\end{equation}}
\newcommand{\bea}{\begin{eqnarray}}
\newcommand{\eea}{\end{eqnarray}}
\def\bra#1{\left\langle #1\right\vert}
\def\ket#1{\left\vert #1\right\rangle}

\def\PLB{{ Phys. Lett.} B}

\def\PRD{{ Phys. Rev.} D}
\begin{document}
{
\tighten
%\preprint {\vbox{
% \hbox{WIS-99/26/July-DPP}
% \hbox{TAUP 2583-99}
% \hbox{hep-ph/9907551} 
% \hbox{ANL-HEP-PR-99-70}
%}}
  
\title {Neutrino oscillations as a ``which-path" experiment}
\author{Harry J. Lipkin\,\thanks{Supported
in part by grant from US-Israel Bi-National Science Foundation
and by the U.S. Department
of Energy, Division of High Energy Physics, Contract W-31-109-ENG-38.}}
\address{ \vbox{\vskip 0.truecm}
  Department of Particle Physics
  Weizmann Institute of Science, Rehovot 76100, Israel \\
\vbox{\vskip 0.truecm}
School of Physics and Astronomy,
Raymond and Beverly Sackler Faculty of Exact Sciences,
Tel Aviv University, Tel Aviv, Israel  \\
\vbox{\vskip 0.truecm}
High Energy Physics Division, Argonne National Laboratory,
Argonne, IL 60439-4815, USA\\
~\\HJL@axp1.hep.anl.gov
\\~\\
}
 
\maketitle
 
\begin{abstract}
 
The role of simple quantum mechanics in understanding neutrino  oscillation
experiments is pointed out by comparison with two-slit and Bragg  scattering
experiments. The importance of considering the beam and the detector as a
correlated quantum system is emphasized. Quantum mechanics alone shows that
the difference observed in the same neutrino detector at Super-Kamiokande
between upward and downward going neutrinos requires the existence of a
neutrino mass  difference. The localization of the source and detector in
space  in the laboratory system for long times leads to an uncertainty in the
momentum but not of the energy of the neutrino and to coherence between states
having different momenta and the same energy and not between states with
different energies. 

\end{abstract}
 
} % end tighten

\section{The Basic Quantum Mechanics of Neutrino Oscillations} 
 
\subsection {Coherence and the momentum uncertainty}

Coherence and interference in neutrino oscillations have been extensively
discussed and clarified
\cite{Dost,QM,NeutHJL,Kayser,GoldS,Pnonexp,MMNIETO,GrossLip,Leo,okun1,pnow98}
but there is still considerable confusion.  The standard textbook neutrino wave
function, a coherent linear combination of states with different energies,
never exists in the real world. Elementary quantum mechanics and quantum 
statistical mechanics tell us that components with different energies in an
initial state are never coherent\cite{Leo}while components with different momenta must
be coherent. The probability must vanish for finding a neutrino source 
outside the tiny region of space where the source is known to exist. 
Any wave packet or density matrix describing the source as a linear
combination  of plane wave momentum eigenstates which exist over all space with
constant  amplitudes must somehow conspire to produce this cancellation outside
the source. This coherence between states having the same energy and different 
momenta  produces  coherence between neutrino states with the same energy and
different masses.

\subsection{Simple Quantum Mechanics and Super-Kamiokande}

Simple quantum mechanics alone, without the full apparatus of the standad
model, shows that the Super-Kamiokande results\cite{SuperK} require the 
existence of two different mass eigenstates for neutrinos.  The energy spectrum
of atmospheric neutrinos cannot change between their source at the top of the
atmosphere and their detection in a detector on earth if  neutrinos are not
absorbed and do not decay en route and any interactions en  route conserve
energy. The momentum spectrum for neutrinos of a given energy is a set of delta
functions, one for each neutrino mass value. If there is only one mass value,
the energy and momentum spectra will be identical for the  upward and downward
going neutrinos incident on the detector and no difference between them can be
observed. The observation of such a  difference\cite{SuperK}  therefore 
indicates that there are at least two different mass eigenstates, and that  the
difference can arise from interference between the waves of states  having
different masses and therefore different momenta if they have the same
energy.     

\subsection {The Static Point Source Approximation} 

If a neutrino is emitted from a point source which
is at a definite position in the laboratory for all time, the neutrino energy 
can be determined precisely by measuring the energy of the source before and
after the neutrino emission. But the point source localization 
introduces an infinite momentum uncertainty. In a realistic case, the source
still remains undisturbed
for a sufficiently long time on the relevant time scale, and its finite size 
is still very much smaller than the wave lengths in space of any neutrino 
oscillation and the distance between the source and the detector. 
Thus the static point source provides  a very good approximation for determining 
which amplitudes are
coherent and which are incoherent. 
Amplitudes describing neutrino states with the same energy and different 
momenta are coherent and must be summed before squaring, while amplitudes
having different energy are incoherent and are squared before summation. 
This is discussed quantitatively below.

\section{The Analog with Two-Slit and Bragg Scattering Experiments} 

The wave-particle duality and quantum mechanics inherent in a neutrino-oscillation
experiment can be clarified  by considering it as a typical ``which-path"
experiment\cite{ADY}. Just as in the two-slit electron diffraction experiment and in
coherent Bragg scattering of photons by a crystal, the neutrino oscillation
experiment describes the emission of a particle from a source and its detection by a
detector separated from the source by a macroscopic distance. There there is no
measurement of the precise path taken by the particle from the source to the
detector. The amplitude at the detector is the coherent sum of the amplitudes from
all allowed paths. 

In the Bragg scattering experiment, the photon may be scattered by any one of
the atoms in the crystal, transfering momentum and energy, but which atom
scattered the photon is not known. In a neutrino oscillation experiment, the
neutrino carrying momentum and energy from the source to detector may be any
one of the allowed neutrino mass eigenstates, but which mass eigenstate carries
this momentum and energy is not known. Here the relevant paths are in
energy-momentum space, rather than configuration space. It is not simple
ignorance which conceals the information on the neutrino mass. Simple ignorance
of which path is taken by a particle does not introduce coherence between
amplitudes. 

Coherence results only from an uncertainty required by quantum mechanics. 
Both in Bragg scattering and neutrino oscillations there
would be no coherence if the energy and momenta of all relevant particles were
measured precisely. The positions both of the atoms in the crystal and of the neutrino
source in the laboratory are known to a precision which produces a sufficiently
large momentum uncertainty to prevent the identification of the scattering atom
or of the neutrino mass. These uncertainties prevent the use of momentum conservation
to distinguish between different possible amplitudes leading to the same final
state at the detector. Because the experimental setup is crucial to the
determination of which amplitudes are coherent, the relevant conditions
determined by the experimental setup  must be  introduced into any calculation
from the beginning. It is thus desirable to work at all times in the laboratory
system, where the source, detector and scattering apparatus are not moving and
the constraints from the uncertainty principle are most simply described. 

\section {Which Path or Witch Craft?}    

Further insight into the physics of Which-Path experiments is given by noting 
the existence of quantum detectors and including the quantum mechanics of the
detector in the analysis of the experiemnt.

\subsection {Classical and Quantum Detectors}

A classical detector in one path of a two-path experiment determines which 
path was taken and destroys all coherence and interference. 
A quantum detector is a quantum system in one path of a which-path experiment.
If a particle passes through its path, it undergoes a transition denoted by
$\ket{D_i}\rightarrow \ket{D_f}$, where  $\ket{D_i}$  and  $\ket{D_f}$ denote
the initial and final states of the detector. 

Consider a simple ``two-slit" which-path experiment in which a quantum
detector  is  introduced\cite{ADY,LipAB} into one path. A particle beam is split
into two paths and the two amplitudes, denoted by $\ket{L(x)}$ and
$\ket{R(x)}$  are then recombined at a point $x$ on a screen.

If no path detector is present 
the wave function and the intensity at the point $x$ are 
\beq{WP11}
 \Psi(x) = \ket{L(x)} + \ket{R(x)}   
 \eeq
\beq{WP12}
I(x) = |\Psi(x)|^2 = |\ket{L(x)}|^2 + |\ket{R(x)}|^2 + 
2 Re [\langle {L(x)} \ket{R(x)}]       
 \eeq
This can be rewritten 
\beq{WP17}
I(x) = |\ket{L(x)}|^2 + |\ket{R(x)}|^2 + 
2 Re [|\langle {L(x)} \ket{R(x)}|\cdot e^{i\theta (x)}] 
 \eeq
\beq{WP18}
 I(x) = |\ket{L(x)}|^2 + |\ket{R(x)}|^2 + 
2 |\langle {L(x)} \ket{R(x)}| \cos \theta (x) 
 \eeq
 where $\theta (x)$ is relative phase of $\ket{L(x)}$ and $\ket{R(x)}$ 

If there is a quantum detector in the ``R" path,
the wave function for the combined system of the particle and the detector and
the intensity observed at $x$ are 
\beq{WP13}
 \Psi(x,D) = \ket{L(x),D_i} + \ket{R(x),D_f}       
 \eeq
\beq{WP14}
 I(x) = |\ket{L(x)}|^2 + |\ket{R(x)}|^2 + 
2 Re [\langle {L(x)} \ket{R(x)} \cdot \langle D_i \ket {D_f}]   
 \eeq

The quantum detector introduces an additional factor in the interference term,
the detector overlap $\langle D_i \ket {D_f} $ . It can also have an additional
phase introduced by the phase of $\langle D_i \ket {D_f} $ 

The Bragg scattering process is an example of a which-path experiment with
many  paths, one for each scattering atom, and a quantum detector in each path.
The detector is the full lattice and each  interference term between two paths
contains two  coherence factors  $\langle D_i \ket {D_f}, $ one for each path,
that depend on the lattice dynamics. The probability $P_{DW}$ that the
scattering is coherent is called the ``Debye-Waller" factor\cite{Mossb,LipAB} and is
just given by

\beq{WP94}
P_{DW} = |\langle D_i \ket {D_f} |^2
 \eeq
   
\subsection {A Simple Toy Model for a Quantum Detector}

Consider the following modification of the toy model of  Stern et al\cite{ADY} 
in which the particle moving in the ``R" path interacts with an 
external spin- one-half object and produces a rotation of this external spin by exactly
$180^o$ about the $z$-axis, while if the particle passes through the ``L" path 
there is no effect.  Then
\beq{WP15}
\ket {D_f} = e^{i\pi s_z} \ket {D_i} = 
e^{i\pi \sigma_z/2} \ket {D_i} 
 \eeq 
\beq{WP16}
\langle D_f \ket {D_i} =
\bra { D_i} e^{i\pi \sigma_z/2} \ket {D_i} 
=\bra { D_i} i\sigma_z \ket {D_i} = i \langle \sigma_z \rangle_i  
 \eeq

The wave function and intensity at the point $x$ on the screen are now

\beq{WP19}
\Psi(x,D) = [\ket{L(x),D_i} + i\langle \sigma_z \rangle_i  \ket{R(x),D_i}] 
 \eeq

\beq{WP20}
I
(x) = |\ket{L(x)}|^2 + |\ket{R(x)}|^2 - 
2 |\langle {L(x)} \ket{R(x)}| \sin \theta (x)]\cdot 
\langle \sigma_z \rangle _i  
 \eeq

The interference term with quantum detector  contains an 
additional factor $ \bra { D_i} \sigma_z \ket {D_i} = 
\langle \sigma_z \rangle _i  $
with an extra $90^o$ phase.
 
   If the spin is initially polarized in the any direction normal to the $z$
axis, then $ \langle \sigma_z \rangle _i  =0$ and there is no interference
between the two paths. One path flips the external spin; the other does not,
and the detector  determines the path.

  If the  spin is initially polarized in the $z$-direction, then 
   $\langle \sigma_z \rangle _i  =\pm 1$  and the rotation
does not change the spin state; it only introduces a phase. There is no 
dephasing, just the addition of a constant relative phase. 

The interesting case  which illustrates the difference between classical and
quantum detectors is when the spin is initially polarized in another 
direction; e.g.  at $45^o$ relative to the $z$-axis in the $x-z$ plane, with
the $z$ and $x$ components both positive. Here $\langle \sigma_z \rangle _i  =
1/2$.

      Classically it is always possible to know the path taken by the particle.
If the spin is rotated the ``R" path has been taken; if the spin is not rotated, 
the ``L" path has been taken. The rotation brings the spin into a direction in the
$x-z$ plane which is still at $45^o$ relative to the $z$ axis and normal to the 
original direction. The $z$-component is still positive but the $x$ component is
negative. 

      This rotation is easily detected classically but not
quantum-mechanically. The initial spin state which is 100\% polarized positive
relative to an axis  at $45^o$  with respect to the $z$ axis with both $x$ and
$z$ components positive is a 50-50 mixture of both positive and negative
polarizations relative to an axis normal to the initial polarization direction
with the $x$ component  negative.

      Thus if we know that the initial spin is polarized as above, and we now 
measure the polarization in the direction of the classically expected final 
polarization, we will indeed find that the final spin is 100\% polarized as 
expected from the classical analysis if the particle went through the path that
interacts with the spin. But if the particle went through the other path and
did not affect the spin at all, the spin is completely unpolarized with respect
to  this new axis, 50\% positive  and 50\% negative. 

      This is thus only a ``partial which path" experiment with partial
dephasing.. The initial and final  states of the spin before and after the
rotation are very different and  distinguishable classically. But
quantum-mechanically they are not orthogonal. The overlap defines a domain
where it is impossible to determine ``which path" and interference will still
be observed.

\section {Detailed Quantum Mechanics of Neutrino Detector}

We now apply the general which-path formalism develped above to a neutrino -
detector system. The wave function for the initial state of neutrino and detector
can be written 
\beq{WP21}
\Psi_i(\nu,D) = 
\sum_{k=1}^{N_\nu} \ket{\nu(E_\nu,m_k,\vec P_k),D_i(E_i)}  
 \eeq
where $N_\nu$ is the number of neutrino mass states, 
$E_\nu$, $m_k$ and $\vec P_k$ denote the neutrino energy, mass and momentum and 
$D_i(E_i)$ is
the initial state of the detector with energy 
$E_i$. 
If the detector is a muon detector the final detector state after neutrino 
absorption is 
\beq{WP23}
 \Psi_f(\mu^\pm,D) = 
\sum_{k=1}^{N_\nu} \ket{\mu^\pm(E_\mu,\vec P_\mu),D^\mp_{kf}(E - E_\mu)}  
 \eeq
where  
$E_\mu$ and $\vec P_\mu$ denote the muon energy and momentum,  $D^\mp_{kf}$ is
the final detector state produced in the ``path $k$"; i.e. after the absorption
of a neutrino with mass  $m_k$ and emission of a $\mu^\pm$, and  
$E=E_\nu + E_i$ is the total energy which is conserved in the transition. 

The transition in the detector occurs on a nucleon, whose co-ordinate is
denoted by by $\vec X$, and involves a charge exchange denoted by the isospin
operator $I_{\mp}$ and a mementum transfer $\vec P_k -\vec P_\mu$.
The detector transition matrix element is therefore given by 
\beq{WP24} 
\bra {D^\mp_{kf}} T^{\mp}\ket {D_i} =  
\bra {D^\mp_{kf}}I_{\mp}e^{i(\vec P_k -\vec P_\mu) \cdot \vec X}  \ket {D_i} 
 \eeq 
 
The overlap between the final detector wave functions after the transitions 
absorbing neutrinos with masses $m_k$ and $m_j$ is then 
 \beq{WP25} 
\langle {D^\mp_{kf}}\ket {D^\mp_{jf}} =  
 \bra {D_i} e^{i(\vec P_j-\vec P_k) \cdot \vec X}  \ket{D_i}  \eeq

If the quantum fluctuations in the position of the active nucleon in the
initial state of the detector are small in comparison with the oscillation wave 
length, $\hbar /(\vec P_j-\vec P_k)$,
\beq{WP26}
|\vec P_j -\vec P_k|^2 \cdot \bra {D_i} |\vec X^2|\ket {D_i} \ll  1      
 \eeq 
\beq{WP27}
\langle {D^\mp_{kf}} \ket {D^\mp_{jf}} \approx  
1 - (1/2)\cdot|\vec P_j -\vec P_k|^2 \cdot \bra {D_i} |\vec X^2|\ket {D_i} 
\approx 1
 \eeq

There is thus effectively a full overlap between the final detector states after
absorption of different mass neutrinos, and a full coherence between the neutrino
states with the same energy and different momenta. 

The total energies of the final muon and detector produced after absorption of
neutrinos with different energies are different. These muon-detector states are thus
orthogonal to one another and there is no coherence between detector states produced
by the absorption of neutrins with different energies. 
 
\section{Conclusions}

\subsection {What we know from simple quantum mechanics} 

Neutrinos propagate from the source to the detector as ordinary Dirac particles
moving freely in space if they are not interacting with matter. They do not get lost
in transit and the relative number of the different mass eigenstates is the same at
the detector as at the source. Only the relative phase between the different mass
eigenstates can change in the propagation from the source to the detector . 

The observation of a difference between upward and downward going atmospheric
neutrinos measured in the same detector can have only two possible explanations.
   
    1. At least two different neutrinos with different masses are emitted
from the source and observed in the detector, and the detector is sensitive to
the relative phases of the waves arriving from neutrinos with different masses.
These relative phases increase monotonically with distance as a well-known
function of the unknown neutrino mass differences, thereby producing
oscillations in the signal observed at the detector as a function of distance.
The experimental results place constraints on the values of the neutrino
mass differences and the couplings of the different neutrino mass states to the
source and the detector (mixing angles in the language of the standard model). 

    2. The neutrinos traveling through the earth do not propagate freely, but 
interact with matter. This is generally known as the MSW effect. These
interactions can change the relative magnitudes as well as the relative phases
of the neutrino mass eigenstates reaching the  detector. They can transfer
momentum, but they conserve energy like a ball  bouncing elastically against
the earth. 

    All these conclusions depend only upon quantum mechanics.

\subsection {What we think we know from the standard model}

In the standard model all the neutrinos observed so far in experiments 
originate in a source from weak  decays or $W$ and $Z$ exchanges and are
detected via $W$ or $Z$ exchange in a detector. The couplings of the three
neutrino mass eigenstates to the three charged leptons and the $W$ is described
by a $3 \times 3$ unitary matrix analogous to the CKM matrix in the quark
sector. These are usually described in terms of mixing angles. 

\subsection {What we don't know and need to determine from experiment}

The masses of the three types of neutrinos and the mixing angles describing
their couplings to the $W$ are completely unknown and are free parameters in
the standard model. We really do not know if the standard model relations
between couplings are really valid and whether new physics beyond the standard
model might influence these relations. However, we emphasize that there is no
justificaton for believing that new physics beyond the standard model can
violate quantum mechanics. Thus the conclusions from quantum mechanics
described above hold even if the standard model is not valid. 

\subsection{Energy-Momentum Kinematics}

We now use the above considerations to specify the relevant scales in neutrino
oscillation experiments. Consider a neutrino emitted from a macroscopic source
whose size is described by a linear dimension $S$, and  detected by a
macroscopic detector at a distance  $D \gg S$ from the source. The knowledge of
the source  position leads to uncertainties in the initial source momentum, the
momentum transfer and the neutrino momentum. 

\beq{WP2} 
\delta p_\nu \approx {{\hbar}\over{S}}
  \eeq

The energy of the source before the emission of the neutrino can be measured in
principle with arbitrary precision. The energy after the  emission can be
measured during the time of flight of the neutrino from source to detector.
This leads to an uncertainty in energy transfer and the neutrino energy
\beq{WP3} \delta E_\nu \approx {{\hbar c}\over{D}} \ll c \delta p_\nu \eeq

The uncertainty in the square of the neutrino mass is then given by 

\beq{WP4} \delta (m_\nu)^2 \cdot c^4=  2 p_\nu \cdot \delta p_\nu \cdot c^2+  2
E_\nu \cdot \delta E_\nu  \approx 2  p_\nu \cdot c^2 \cdot
\left({{\hbar}\over{S}} + {{\hbar}\over{D}}\right)    \approx 2 p_\nu \cdot c^2
\cdot {{\hbar}\over{S}}  \eeq Interference effects can be observed at the
detector between  the contributions from neutrino states with different masses
if the squared mass difference is less than this value (\ref{WP4}). The
uncertainty in neutrino mass arises from the uncertainty in the neutrino
momentum. Eq. (\ref{WP3}) shows that the uncertainty in the neutrino  energy is
negligible. Thus any coherence observed at the detector between  amplitudes
from neutrinos with different masses must come from states with the same energy
and different momenta. 

The relative phase between two neutrino waves with the same energy, masses
$m_1$ and $m_2$ and  momenta $p_1$ and $p_2$ changes in traversing a  distance
D by the amount

\beq{WP5}
\delta \phi(D)= (p_1 - p_2)\cdot{{D}\over{\hbar}} =
{{(p_1^2 - p_2^2)}\over{(p_1 + p_2)}} \cdot {{D}\over{\hbar}} \approx
{{(m_2^2 - m_1^2)\cdot c^4}\over{2p_\nu}}\cdot {{D}\over{\hbar}} 
\eeq

For this phase to be of order unity and give rise to observed neutrino
oscillations, 

\beq{WP6}
m_2^2 - m_1^2 \approx {{2 \hbar p_\nu}\over{D c^4 }} \ll 
 {{2 \hbar p_\nu}\over{S c^4 }} \approx 
\delta (m_\nu)^2 
\eeq
 This squared-mass difference (\ref{WP6}) is much less than
the lower limit on detectable mass difference imposed by the uncertainty
condition (\ref{WP4}). The momentum difference between the different mass
eigenstates having the same energy is much smaller than the momentum
uncertainty produced by the localization of the source. 
Thus the neutrino mass difference needed to produce oscillations with wave
lengths of the order of the source-detector distance $D$ cannot be detected in
any experiment in which the distance from the source to the detector is much
larger than the size of the source.

The wave length of the neutrino oscillations is given directly by eq. 
(\ref{WP5}).

For an example showing characteristic numbers, the neutrino momentum from a
pion decay at rest is $\approx 30$ MeV/c or $3 \times 10^7$ ev/c. If there are
two neutrino masses of 1 and 2 ev. their momentum difference if they have the
same energy is 

\beq{WP7}
p_1 - p_2 \approx   {{(m_2^2 - m_1^2)c^2}\over p} = 10^{-7} \rm{ev/c}
\eeq
Since $\hbar c \approx 2 \times 10^{-7} $ ev $\times$ meters, the oscillation
wave length will be of order 2 meters and knowing the source position with a
precision of more than two meters will prevent the measurement of this 
momentum difference. If the two neutrino masses are 0.1 and 0.2 ev. these 
numbers scale by a factor of 100 and $p_1 - p_2 \approx 10^{-9} $ ev/c and the
oscillation wave length is 200 meters.  

This effectively says it all for neutrino propagation in free space between a
source and detector whose size and distance satisfy the condition 
that the distance between source and detector is much greater than the size of
the source. The point source approximation is good. Except for the case of
matter-induced oscillations and the MSW effect or for the case of propagation
through external fields there is no need to engage in more complicated
descriptions to obtain the desired results.

\section{acknowledgments}

It is a pleasure to thank Maury Goodman, Yuval Grossman,
Boris Kayser, Lev Okun and Leo Stodolsky for helpful discussions and
comments.

\end{document}